# Proton and Electron Irradiations of CH$_4$:H$_2$O Mixed Ices

Duncan V. Mifsud[1,2,*], Péter Herczku[2], Béla Sulik[2], Zoltán Juhász[2], István Vajda[2], István Rajta[2], Sergio Ioppolo[3,4], Nigel J. Mason[1,2], Giovanni Strazzulla[5], and Zuzana Kaňuchová[6,*]

1. Centre for Astrophysics and Planetary Science, School of Physics and Astronomy, University of Kent, Canterbury CT2 7NH, United Kingdom
2. Institute for Nuclear Research (Atomki), Debrecen H-4026, Hungary
3. Centre for Interstellar Catalysis (InterCat), Department of Physics and Astronomy, University of Aarhus, 8000 Aarhus, Denmark
4. School of Electronic Engineering and Computer Science, Queen Mary University of London, London E1 4NS, United Kingdom
5. INAF Osservatorio Astrofisico di Catania, I-95123 Catania, Italy
6. Astronomical Institute, Slovak Academy of Sciences, Tatranská Lomnica SK-059 60, Slovak Republic
* Correspondence: dm618@kent.ac.uk (DVM), zkanuch@ta3.sk (ZK)

**Abstract:** The organic chemistry occurring in interstellar environments may lead to the production of complex molecules that are relevant to the emergence of life. Therefore, in order to understand the origins of life itself it is necessary to probe the chemistry of carbon-bearing molecules under conditions that simulate interstellar space. Several of these regions, such as dense molecular cores, are exposed to ionizing radiation in the form of galactic cosmic rays which may act as an important driver of molecular destruction and synthesis. In this paper, we report the results of a comparative and systematic study of the irradiation of CH$_4$:H$_2$O ice mixtures by 1 MeV protons and 2 keV electrons at 20 K. We demonstrate that our irradiations result in the formation of a number of new products, including both simple and complex daughter molecules such as C$_2$H$_6$, C$_3$H$_8$, C$_2$H$_2$, CH$_3$OH, CO, CO$_2$, and probably also H$_2$CO. A comparison of the different irradiation regimes has also revealed that proton irradiation resulted in a greater abundance of radiolytic daughter molecules compared to electron irradiation, despite a lower radiation dose having been administered. These results are important in the context of the radiation astrochemistry occurring within the molecular cores of dense interstellar clouds, as well as on outer Solar System objects.

## 1. Introduction

A comprehensive understanding of the chemistry of carbon-bearing molecules in space is necessary in order to quantitatively study the extent of organic chemistry in extraterrestrial environments, including reactions that may lead to the formation of complex organic molecules of inherent interest to biology and biochemistry. A number of carbon-bearing molecules have been detected in interstellar space, ranging in structural complexity from methylidyne radicals, CH to fullerenes such as buckminsterfullerene, C$_{60}$ and rugbyballene, C$_{70}$ [1]. Methane, CH$_4$ is the simplest saturated organic molecule which is known to exist in the interstellar medium, having originally been observed in both the solid and gas phases towards the young stellar object NGC 7538 IRS 9 [2].

The initial observations of interstellar CH$_4$ found that a large fraction was locked away in interstellar icy grain mantles, thus suggesting a synthesis pathway that relied upon reactions in the solid phase. Indeed, recent experimental studies have demonstrated that carbon and hydrogen atom addition reactions on the surfaces of interstellar dust grains are able to produce CH$_4$, and that the rate of formation is twice as high when CH$_4$ is formed in an ice that is rich in solid water, H$_2$O [3]. Such experimental results complement the previously observed correlations between $CH_4$ and $H_2O$ ices in interstellar icy grain mantles observed towards low-mass young stellar objects [4]. $CH_4$ has also been observed in comets [5-7], implying the possible survival of interstellar $CH_4$ during the processes of stellar and planetary system formation. Cometary impacts could indeed have played a key role in the delivery of $CH_4$ to various planetary-like bodies in the Solar System, such as Titan or Pluto [8,9].

Interstellar and outer Solar System environments are characterized by chemistry driven by ionizing radiation mediated by galactic cosmic rays or the solar wind. As such, it is conceivable that the radiation chemistry of $CH_4$ in interstellar or outer Solar System ices could be a driver towards the formation of complex organic molecules in these environments. Such radiation chemistry is known to be driven by the release of several thousand low-energy secondary electrons along the track of the irradiating particle [10,11]. Previous laboratory studies have investigated the radiation chemistry of pure $CH_4$ astrophysical ice analogues using keV-MeV ions and electrons [12-16], and have demonstrated the formation of various daughter hydrocarbons, including: ethane, $C_2H_6$; propane, $C_3H_8$; ethene, $C_2H_4$; and ethyne, $C_2H_2$ as well as several radical species sourced from their monodehydrogenation. When mixed with other molecules, irradiated $CH_4$-containing ices yield even more complex molecules: electron-irradiated mixtures with phosphine, $PH_3$ yield methylphosphanes as large as $CH_3P_8H_9$ [17], while the formation of higher-order carboxylic acids such as decanoic acid, $C_9H_{19}COOH$ has been reported in irradiated mixtures with carbon dioxide, $CO_2$ [18]. The inclusion of nitrogen-bearing molecules, such as ammonia, $NH_3$ into electron-irradiated $CH_4:CO_2$ mixed ices has also been shown to result in the formation of glycine [19]; a proteinogenic amino acid of direct relevance to biology.

Of interest to this present study is the radiation chemistry occurring in $CH_4:H_2O$ mixed ices due to the ubiquity of $H_2O$ ices within astronomical environments [20], as well as the apparent association of $CH_4$ formation with $H_2O$-rich environments [2,3]. This radiation chemistry has been considered by previous experimental studies [21-24] which have shown that alcohols, higher hydrocarbons, and methanal, $H_2CO$ are among the readily produced closed-shell daughter molecules. For instance, the study of Wada *et al.* [22] considered the low-energy electron irradiation of $CH_4:H_2O$ ices at 10 K and found that methanol, $CH_3OH$ is the major daughter molecule of this process and is formed by the combination of hydroxyl, OH and methyl, $CH_3$ radicals or alternatively by the insertion reaction between methylene radicals, $CH_2$ and $H_2O$. In their study on the irradiation of $CH_4:H_2O$ ices by 40 MeV nickel ions at 15 K, Mejía *et al.* [23] were able to additionally demonstrate the formation of ethanol, $C_2H_5OH$; ethanal, $CH_3CHO$; and methanoic acid, HCOOH along with simpler molecules such as carbon monoxide, CO; $CO_2$, and $H_2CO$. Similar results were reported by Krim and Jonusas [24], who also reported the formation of a number of ether molecules as well as glycolaldehyde, $HOCH_2CHO$ after the irradiation of a $CH_4:H_2O$ ice by vacuum-ultraviolet photons at 3 K.

To the best of our knowledge, however, no previous study has comparatively investigated the radiation chemistry induced in $CH_4:H_2O$ ices by different types of energetic processing despite the evident astrochemical importance of such ices. Understanding the potential differences in the radiation chemistries induced by different processing types is important since different astronomical environments are subject to different combinations of ion, electron, and ultraviolet photon irradiations and the predominance of one type of energetic processing over another may lead to the production of distinctive daughter molecules in varying abundances. Indeed, the work of Moore and Hudson [25] demonstrated that several daughter molecules produced after the proton irradiation of $CH_4$ ices mixed with either CO or molecular nitrogen, $N_2$ are not produced after similar irradiations using ultraviolet photons. Baratta *et al.* [12] also performed a comparative study in which they investigated the irradiation of pure $CH_4$ ices by 30 keV $He^+$ and 10.2 eV photons and found that similar administered doses resulted in a greater extent of $CH_4$ destruction and daughter molecule formation in the case of the $He^+$ irradiation. Indeed, they observed that $C_2H_6$, itself formed from the dissociation of the $CH_4$ parent species, began to decay after a dose of about 30 eV per 16 u had been administered *via* He$^+$ irradiation; such an observation was not made during irradiation with 10.2 eV photons.

In this paper, we present the results of a systematic and comparative study looking into the irradiation of CH$_4$:H$_2$O mixed ices using 1 MeV protons and 2 keV electrons so as to characterize the differences in daughter molecule formation and abundances between the different types of energetic processing. The experimental methodology used to carry out this study is described in Section 2, while our results are presented in Section 3 and are discussed in the context of their applicability to astrochemistry. A summative conclusion is given in Section 4.

## 2. Methodology

The experimental work was carried out using the Ice Chamber for Astrophysics-Astrochemistry (ICA) at the Institute for Nuclear Research (Atomki) in Debrecen, Hungary. This set-up has been described in great detail in previous publications [26,27] and so in this paper we limit ourselves to only the most salient details. The ICA is an ultrahigh-vacuum compatible steel chamber with a base pressure of a few 10$^{-9}$ mbar which is achieved through the combined action of a turbomolecular pump and a dry rough pump. In the center of the chamber is a gold-coated oxygen-free high-conductivity copper sample holder into which a maximum of four zinc selenide, ZnSe substrates may be mounted (Figure 1). The sample holder and the substrates may be cooled to 20 K through contact with the cold finger of a closed-cycle helium cryostat, although an operational temperature range of 20-300 K is available.

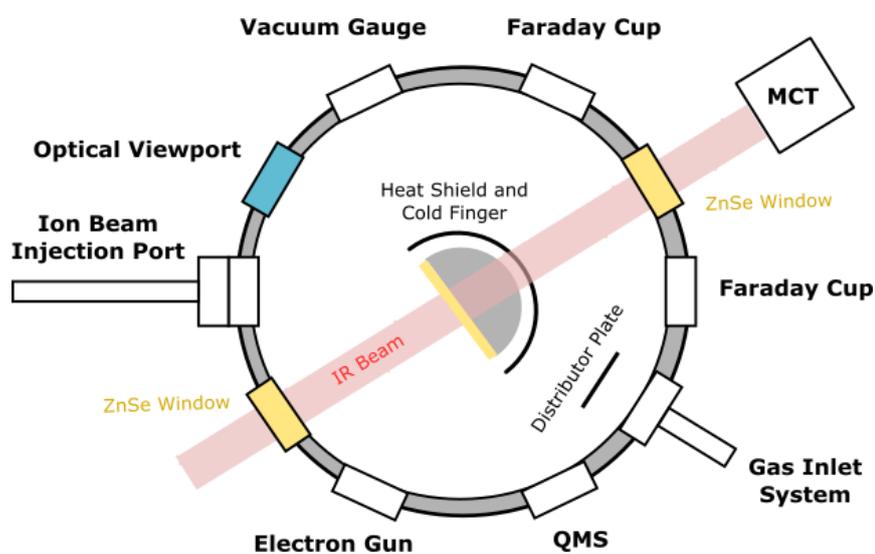

**Figure 1.** Top-view schematic diagram of the ICA set-up. Figure reproduced from Mifsud *et al.* [27] with the kind permission of the European Physical Journal (EPJ).

CH$_4$:H$_2$O astrophysical ice analogues were prepared on the ZnSe substrates by background deposition of a gas mixture at 20 K. The gas mixture was first prepared in a mixing line with the desired stoichiometric ratio (CH$_4$:H$_2$O = 5:2) by introducing CH$_4$ gas (Linde; 99.5% purity) and H$_2$O vapor (from a high purity liquid sample that had been de-gassed by performing several freeze-pump-thaw cycles using liquid nitrogen). The gas mixture was then dosed into the main chamber at a pressure of a few 10$^{-6}$ mbar until ices of approximately 0.25 μm thickness were prepared. The deposition and growth of the ices were monitored *in situ* using Fourier-transform mid-infrared transmission absorption spectroscopy (resolution = 1 cm$^{-1}$; spectral range = 4000-650 cm$^{-1}$) and an external MCT detector. Once deposited, a pre-irradiation mid-infrared spectrum was acquired which allowed us to quantify the column densities (*N*, molecules cm$^{-2}$) of the deposited molecules *via* Eq. 1:

$$N = \frac{P \ln(10)}{A_\nu} \quad (1)$$

where $P$ is the area of a characteristic absorption band (cm$^{-1}$) of the species of interest and $A_\nu$ is its associated integrated band strength constant (cm molecule$^{-1}$). The factor of ln(10) allows for the relation of $P$ (which is measured from spectra on an absorbance scale) to $A_\nu$ (which is measured on an optical depth scale). Deposited CH$_4$:H$_2$O ices were then exposed either to a 1 MeV proton beam delivered by a Tandetron particle accelerator [28,29] or to a 2 keV electron beam delivered by an electron gun mounted directly to one of the side ports of the ICA [27]. The homogeneities, profiles, and currents of the proton and electron beams were respectively assessed using the methods described in Herczku *et al.* [26] and Mifsud *et al.* [27] prior to commencing the irradiation of the ices. Beam currents were measured to be 290 nA in the case of the proton beam and 5.7 μA in the case of the electron beam. Additional mid-infrared spectra were collected throughout the irradiation process so as to identify the formation of daughter molecules. A summary of the experiments conducted in this study (including information on the physical parameters required to perform quantitative analysis) is provided in Table 1.

**Table 1.** Summary of the experiments performed in this study.

| Parameter | Experiments | |
| --- | --- | --- |
|  | 1 | 2 |
| CH$_4$ column density (10$^{17}$ cm$^{-2}$)[a] | 3.24 | 3.91 |
| H$_2$O column density (10$^{17}$ cm$^{-2}$)[b] | 1.29 | 1.59 |
| Projectile | 1 MeV proton | 2 keV electron |
| Projectile stopping power (eV Å$^{-1}$)[c] | 2.1 | 0.9 |
| Fluence delivered (cm$^{-2}$) | 4.95×10$^{15}$ | 4.16×10$^{16}$ |
| Dose delivered (eV per 16 u) | 46.30 | 168.42 |
| Irradiation temperature (K) | 20 | 20 |
| Ice thickness (μm) | 0.22 | 0.27 |

[a] Calculated using the area of the $\nu_4$ mode at ~1300 cm$^{-1}$ assuming $A_\nu$ = 1.03×10$^{-17}$ cm molecule$^{-1}$ [30].
[b] Calculated using the area of the $\nu_s$ mode at ~3370 cm$^{-1}$ assuming $A_\nu$ = 1.50×10$^{-16}$ cm molecule$^{-1}$ [31].
[c] Calculated using the SRIM software [32].

## 3. Results and Discussion

The mid-infrared spectra of the CH$_4$:H$_2$O ices after their preparation at 20 K as well as during their processing by 1 MeV proton and 2 keV electron irradiation are given in Figure 2. Upon preparation, various mid-infrared absorption features are immediately apparent, such as the $\nu_4$ (1300 cm$^{-1}$), $\nu_2 + \nu_4$ (2816 cm$^{-1}$), $\nu_1$ (2901 cm$^{-1}$) and $\nu_3$ (3011 cm$^{-1}$) modes of CH$_4$ [16], as well as the $\nu_L$ (713 cm$^{-1}$), $\nu_2$ (1634 cm$^{-1}$), and $\nu_s$ (3370 cm$^{-1}$) modes of H$_2$O [33]. Of particular interest is the single, strong absorption band present at around 3616 cm$^{-1}$ which is attributed to the hydroxyl dangling bond vibration of H$_2$O [33-36]. In pure H$_2$O ices, the dangling bond vibration typically presents as two weak absorption features; however, when mixed with CH$_4$ these dangling bond absorption bands merge into one significantly stronger band which is red-shifted to lower wavenumbers. This phenomenon has been studied in detail by Gálvez *et al.* [37] and Herrero *et al.* [38], who demonstrated that this could be ascribed to the interaction of CH$_4$ with microporous H$_2$O and that the intensity of the band increases with increasing CH$_4$ content in the mixed ice. Given that our ice is rich in CH$_4$ (CH$_4$:H$_2$O = 5:2), it is therefore not surprising that the hydroxyl dangling bond absorption band is one of the strongest features in the acquired mid-infrared spectra (Figure 2). The interaction of CH$_4$ with microporous H$_2$O is also responsible for the appearance of the symmetry-forbidden $\nu_1$ mode of CH$_4$, which is present in our spectra due to the distortion of the cubic structure of small crystallites resulting in a breakdown of the symmetry of the CH$_4$ molecule [38].

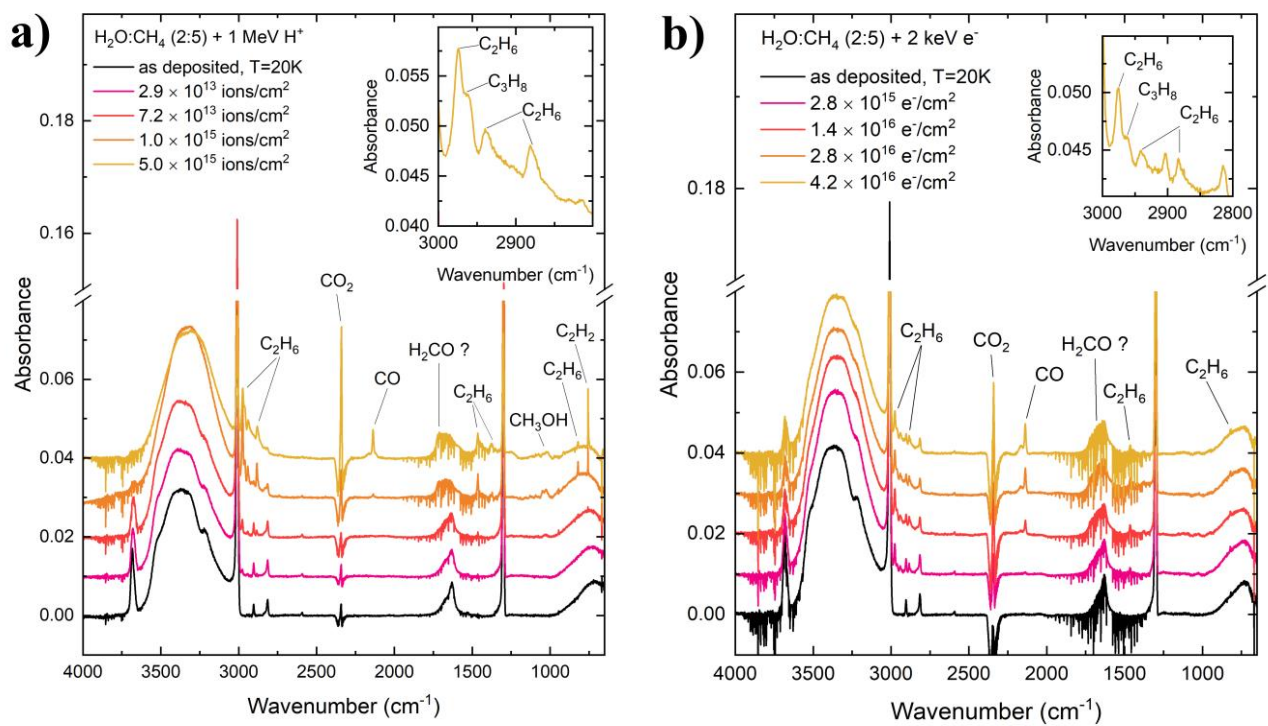

**Figure 2.** Mid-infrared absorption spectra of CH$_4$:H$_2$O (= 5:2) mixed ices upon preparation at 20 K and during irradiative processing using: a) 1 MeV protons, and b) 2 keV electrons. The absorption bands of various daughter molecules have been labelled. Note that the fine structures and negative absorptions in the regions of the H$_2$O and CO$_2$ bands are due to variations in the concentrations of atmospheric gas-phase H$_2$O and CO$_2$ in the pathlength leading to the detector, which is not maintained under vacuum.

The processing of CH$_4$:H$_2$O mixed ices results in the appearance of a number of new absorption bands due to the formation of various daughter molecules as a result of the radiation chemistry occurring within the ices. The most likely first steps leading to the formation of daughter molecules is the radiation-induced dehydrogenation of CH$_4$ and H$_2$O to respectively yield CH$_3$ and OH radicals [21-24].

$$CH_4 \rightarrow CH_3 + H \qquad (2)$$

$$H_2O \rightarrow OH + H \qquad (3)$$

Hydrogen atoms and OH radicals are likely to be very mobile within the ice structure and may thus partake in various hydrogenation or hydrogen abstraction reactions, including the further dissociation of CH$_4$ to CH$_3$ radicals, or the combination reactions leading to the formation of molecular hydrogen, H$_2$ and hydrogen peroxide, H$_2$O$_2$. Neither of these latter two species were detected in our experiments. The non-detection of H$_2$ is ascribed to its lack of a permanent dipole moment which makes it infrared-inactive, as well as its volatility which allows it to efficiently escape from the ice *via* sublimation to the gas-phase; while the non-detection of H$_2$O$_2$ is likely due to its mid-infrared absorption bands being obscured by those of hydrocarbon species which are more abundant within the ice.

$$H \text{ (or OH)} + CH_4 \rightarrow H_2 \text{ (or H}_2\text{O)} + CH_3 \qquad (4)$$

$$H + H \rightarrow H_2 \qquad (5)$$

$$OH + OH \rightarrow H_2O_2 \qquad (6)$$

Given the preponderance of $CH_4$ within our ices, it is most likely that radicals derived from the radiolysis of this species would come into contact with one another. For instance, the combination of two $CH_3$ radicals would yield $C_2H_6$ as a result of the formation of a single bond between the two carbon atoms. Indeed, mid-infrared absorption bands attributable to the $\nu_{12}$ (822 cm$^{-1}$), $\nu_6$ (1375 cm$^{-1}$), $\nu_{11}$ (1464 cm$^{-1}$), $\nu_5$ (2883 cm$^{-1}$), $\nu_8 + \nu_{11}$ (2941 cm$^{-1}$) and $\nu_{10}$ (2975 cm$^{-1}$) modes of $C_2H_6$ [16] were observed after both proton and electron irradiations of the ice mixtures (Figure 2).

$$CH_3 + CH_3 \rightarrow C_2H_6 \qquad (7)$$

The formation of $C_2H_6$ is of inherent astrochemical interest since it may be classified as a complex organic molecule (the astrochemical definition of a complex molecule includes all those with six or more constituent atoms). To the best of our knowledge, $C_2H_6$ has yet to be detected in the interstellar medium [1], although it has been found in various icy outer Solar System objects. Solid $C_2H_6$ has been observed on the surface of Pluto [39] while liquid $C_2H_6$ is thought to be an important component of oceanic and lacustrine geochemical systems on Titan [40,41]. Our results conclusively show that $C_2H_6$ production from $CH_4$ radiolysis is not inhibited by the presence of $H_2O$ in the ice mixture, and thus it may conceivably form by solid-phase radiation chemistry in realistic interstellar icy grain mantles rich in both species [3,4]. Indeed, observational studies of comet C/1996 B2 Hyakutake revealed an abundance of $C_2H_6$ comparable to that of $CH_4$, implying that the comet did not originate in a thermochemically equilibrated astronomical environment [42]. Since the production of $C_2H_6$ by gas-phase ion-molecule reactions is energetically forbidden [43], this observation strongly points to a solid-phase production route towards $C_2H_6$ in dense interstellar cloud cores which is consistent with our experimental results.

The presence of $C_2H_6$ within the ice opens further radiolytic channels towards the formation of higher hydrocarbons: $C_3H_8$ is another daughter species formed after the irradiation of our $CH_4:H_2O$ mixed ices (Figure 2). The presence of $C_3H_8$ in the irradiated mixed ices is likely the result of a two-step process first involving the dehydrogenation of $C_2H_6$ to yield ethyl radicals, $C_2H_5$ followed by the addition of $CH_3$ to yield the observed product.

$$C_2H_6 + H \text{ (or OH)} \rightarrow C_2H_5 + H_2 \text{ (or } H_2O) \qquad (8)$$

$$C_2H_5 + CH_3 \rightarrow C_3H_8 \qquad (9)$$

Furthermore, $C_2H_6$ may partake in successive dehydrogenation reactions leading to the formation of alkenes and alkynes: $C_2H_2$ was efficiently formed after the 1 MeV proton irradiation of our $CH_4:H_2O$ mixed ice, as indicated by the intensity of its $\nu_5$ mode at 757 cm$^{-1}$ [44] (Figure 2). The formation of $C_2H_2$ from $C_2H_6$ is a multi-step process involving four dehydrogenation reactions, with $C_2H_4$ being formed in the process after two hydrogen abstractions. Interestingly, no $C_2H_2$ was detected after the irradiation of our mixed ice using 2 keV electrons (Figure 2). This observation may be the result of the lower molecular destruction and formation cross-sections associated with projectile electrons compared with protons [45,46], and constitutes a major difference in the irradiation experiments considered in this paper. It is also worth noting that $C_2H_4$ was not detected in either the proton or the electron irradiated ice, although we speculate this may have been due to the absorption bands of this molecule having been obscured by the stronger absorption bands of the parent $CH_4$ and $H_2O$ species [16].

$$C_2H_6 + 2\,H \text{ (or 2 OH)} \rightarrow C_2H_4 + 2\,H_2 \text{ (or 2 } H_2O) \qquad (10)$$

$$C_2H_4 + 2\,H \text{ (or 2 OH)} \rightarrow C_2H_2 + 2\,H_2 \text{ (or 2 } H_2O) \qquad (11)$$

Another molecule for which there exists a significant discrepancy in the comparative abundance produced after proton and electron irradiation of the $CH_4:H_2O$ ices is $CH_3OH$, with more of this molecule being formed after proton irradiation compared to electron irradiation (Figures 2 and 3). The formation of $CH_3OH$ may have occurred as a result of a number of different radiolytic pathways. One such pathway is the direct combination of $CH_3$ and OH radicals [47].

$$OH + CH_3 \rightarrow CH_3OH \qquad (12)$$

However, further dehydrogenations of the nascent $CH_3$ radical may yield other reactive carbon-based radicals, including $CH_2$ and CH radicals. The reaction between $CH_2$ and OH, followed by a hydrogenation of the carbon atom could also yield $CH_3OH$.

$$CH_2 + OH \rightarrow CH_2OH \qquad (13)$$

$$CH_2OH + H \rightarrow CH_3OH \qquad (14)$$

Wada *et al.* [22] also provided evidence for the insertion of $CH_2$ into the $H_2O$ molecular structure as an alternative route towards $CH_3OH$ formation.

$$CH_2 + H_2O \rightarrow CH_3OH \qquad (15)$$

The necessary precursor radicals in the reactions leading to the formation of $CH_3OH$ highlighted above have all been observed in experimental studies [24], and thus all likely contribute to its formation in our work to at least some extent.

Yet another formation pathway for $CH_3OH$ in our irradiated ices is the step-wise hydrogenation of CO to sequentially yield HCO, $H_2CO$, $CH_3O$ (or $H_2COH$), and finally $CH_3OH$ [48-50]. The contribution of such a reaction sequence to $CH_3OH$ formation in our ices necessitates the presence of CO, which was observed in our irradiated ices. Although the specific mechanism for CO formation as a result of the irradiation of $CH_4$:$H_2O$ mixed ices is not known for certain, it is possible to suggest that carbon-rich radicals such as CH or even possibly carbon atoms produced as a result of the repeated dehydrogenation of $CH_4$ may have been able to react directly with $H_2O$ to yield CO. Although speculative, such a suggestion is in agreement with previous works that have demonstrated the formation of both CO and $CO_2$ as a result of the irradiation of $H_2O$ ice deposited on top of carbonaceous residues and hydrogenated carbon grains [51-53].

It should be noted that the sequential hydrogenation of CO should also yield $H_2CO$, another closed-shell daughter molecule. This species is usually detected by the observation of its intense $\nu_2$ mode at about 1725 cm$^{-1}$ [31]. However, in our ices, this detection is made somewhat more complicated by two factors: firstly, this band would at least partially coincide with the broad $\nu_2$ mode of $H_2O$. Secondly, the appearance of many fine absorption features in this region of the spectrum due to gas-phase $H_2O$ absorption in the pathlength near the spectroscopic detector makes deconvolution of any possible composite absorption band to its individual contributors practically impossible. However, by looking more closely at Figure 2, it is possible to note that the broad band attributed to the $\nu_2$ mode of $H_2O$ in the pre-irradiation spectra appears to broaden further and blue-shift to higher wavenumbers during irradiation. This could be interpreted as being the result of the appearance of the $H_2CO$ $\nu_2$ mode. However, in the absence of confirmatory spectroscopic deconvolution analyses, we limit ourselves to only tentatively detecting $H_2CO$ in the ices (hence the '$H_2CO$ ?' label in Figure 2).

The final molecule to be spectroscopically detected in our irradiation experiments is $CO_2$ (Figure 2). The formation of $CO_2$ likely proceeded *via* a combination of three radiation-driven processes. Firstly, $CO_2$ formation may occur as a result of the reaction of carbon-rich radicals or carbon atoms with $H_2O$ ice, analogously to the process described previously for CO formation [51-53]. $CO_2$ is also a known dissociation product of $CH_3OH$ [26,27,54-58], and so the radiolytic destruction of this latter species may contribute somewhat to the formation of $CO_2$. Lastly, and possibly most significantly, the formation of $CO_2$ may occur as a result of the reaction of CO with OH radicals *via* the production of the HO–CO intermediate [59,60]. Conversely, the reaction between CO and any oxygen atoms yielded from the radiolytic destruction of some oxygen-containing precursor species is not anticipated to contribute to the presence of $CO_2$ in our ices due to the known high activation energy barriers associated with the reaction [61-63].

$$CO + OH \rightarrow HO\text{--}CO \rightarrow CO_2 + H \tag{16}$$

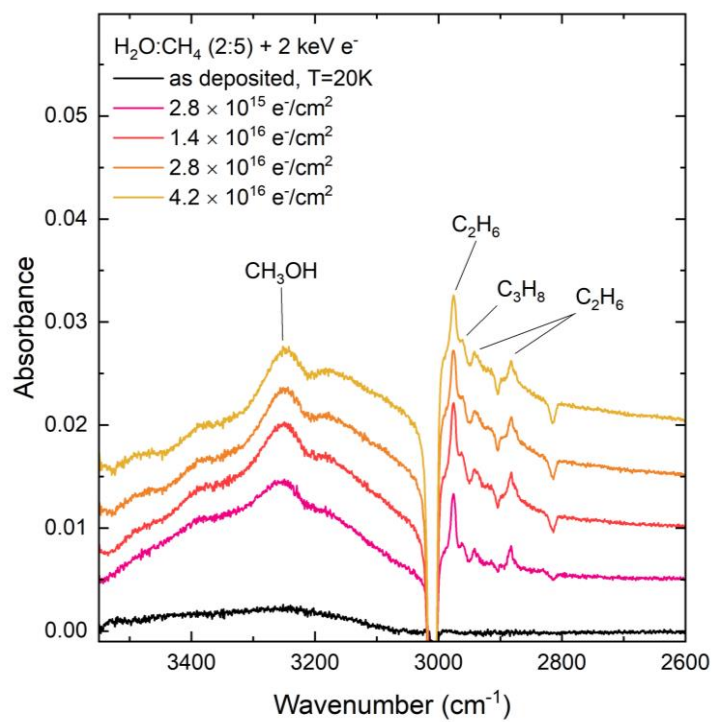

**Figure 3.** Mid-infrared absorption spectra of a $CH_4:H_2O$ (= 5:2) mixed ice upon preparation at 20 K and during irradiative processing using 2 keV electrons. Note that spectra acquired during irradiation are so-called 'difference spectra' yielded after the subtraction of the 'as deposited' spectrum. The spectra are focused on the 3500-2600 cm$^{-1}$ region so as to highlight the formation of $CH_3OH$ as a result of the electron irradiation of the $CH_4:H_2O$ mixed ice.

In order to better quantify the differences in the radiation chemistry of our $CH_4:H_2O$ ices mediated by 1 MeV protons or 2 keV electrons, we have compared the column density evolution as a function of radiation dose for two radiolytic daughter molecules: CO and $C_2H_6$ (Figure 4). By more closely examining the evolution of the column densities of these molecules in both the proton and the electron irradiated ices, it is possible to note that the overall qualitative trend is similar: the column density of $C_2H_6$ begins to increase until it peaks at a given dose (about 19 and 55 eV per 16 u in the case of proton and electron irradiation, respectively), after which it slowly decreases. Conversely, the column density of CO is observed to continually increase with higher radiation doses, with no evidence of peak abundance having yet been reached in either irradiated ice. Interestingly, extrapolation of the proton irradiation data plotted in Figure 4 suggests that the column densities of the $C_2H_6$ and CO daughter molecules are equal at approximately the same dose (about 63 eV per 16 u) in both the proton and the electron irradiated ice.

More generally, it is possible to note that the column densities of daughter molecules measured after proton irradiation are significantly greater than those measured after electron irradiation (Figure 4), despite similar initial (i.e., pre-irradiative) $CH_4$ and $H_2O$ column densities in each case. Once again, we speculate that this is the result of the larger cross-sections associated with molecular formation and destruction induced by proton irradiation compared to electron irradiation [45,46]. We note that this reason was not invoked by Baratta *et al.* [12], who investigated the comparative processing of pure $CH_4$ ice using 30 keV He$^+$ and 10.2 eV ultraviolet photons. In their study, Baratta *et al.* [12] noted that, at low radiation doses of less than 20 eV per 16 u, the decay of CH$_4$ and the concomitant production of C$_2$H$_6$ followed very similar trends in both the He$^+$ and the photon irradiation experiments, both qualitatively and quantitatively. At higher doses, however, the decay of CH$_4$ proved to be more rapid in the case of He$^+$ irradiation. Furthermore, the column density of C$_2$H$_6$ eventually peaked and subsequently began to decay during He$^+$ irradiation, whereas it eventually plateaued during ultraviolet photon irradiation.

The differences observed by Baratta *et al.* [12] were ascribed to the dependence of the photochemistry induced by the ultraviolet photons on the optical properties of the ice which would have undoubtedly changed as a result of its changing chemical composition, thus attenuating many of the incident photons at the most heavily processed regions closest to the surface of the ice. This contrasts with our study, where the radiation chemistry induced by both the projectile protons and electrons was independent of the optical properties of the ice, and both projectile types were able to access practically the entirety of the deposited ice. As such, differences in the total cross-sections of each radiation process seem to be the most parsimonious explanation for the observed differences in the column densities of the radiolytic daughter molecules (Figure 4).

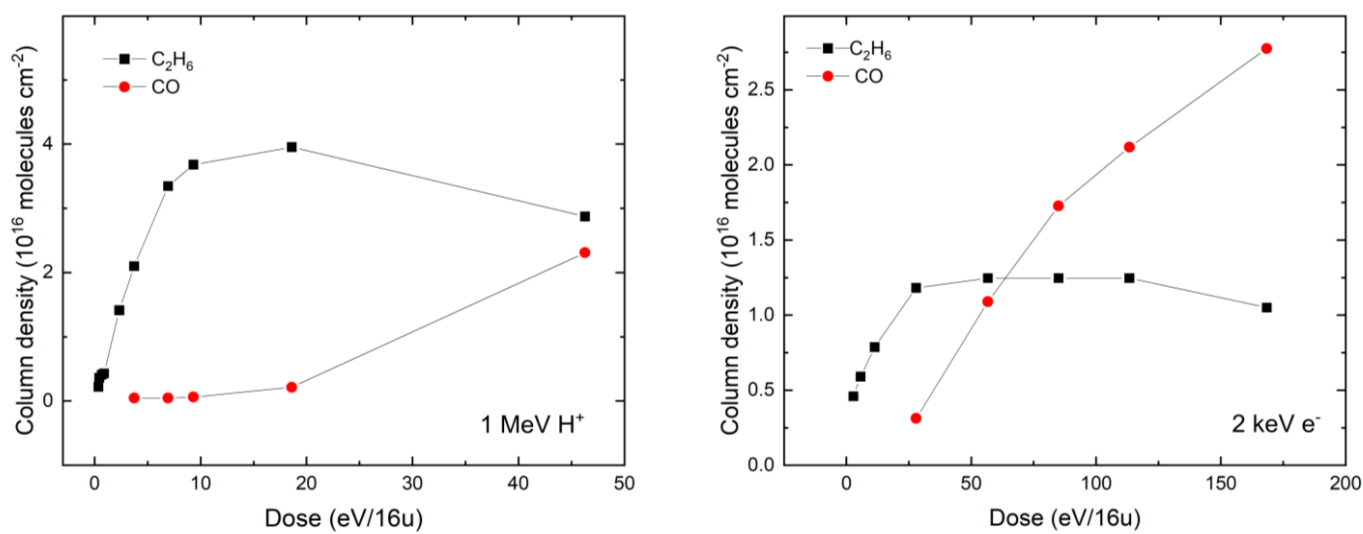

**Figure 4.** Dose evolution of the column densities of CO and C$_2$H$_6$ radiolytic daughter molecules during 1 MeV proton irradiation (*left panel*) and 2 keV electron irradiation (*right panel*) of a CH$_4$:H$_2$O (= 5:2) mixed ice. Column densities were calculated using integrated band strength constants of 3.51×10$^{-18}$ and 1.12×10$^{-17}$ cm molecule$^{-1}$ for the C$_2$H$_6$ absorption band at 2883 cm$^{-1}$ and the CO absorption band at 2139 cm$^{-1}$, respectively [31,64].

## 4. Conclusions

In this paper, we have reported the results of 1 MeV proton and 2 keV electron irradiations of CH$_4$:H$_2$O (= 5:2) mixed ices representative of those that may conceivably exist within icy grain mantles in dense interstellar clouds. Our results have shown that both irradiations result in the formation of many new molecules, some of which are considered to be complex organic molecules in the field of astrochemistry. Among these radiolytic daughter molecules were C$_2$H$_6$, C$_3$H$_8$, CH$_3$OH, CO, CO$_2$, and probably H$_2$CO. Proton irradiation of the ice mixture additionally resulted in the production of C$_2$H$_2$, which was not observed after electron irradiation of the ice despite a significantly higher radiation dose being administered during electron irradiation. Our results have further demonstrated that the abundances of these daughter species are greater after proton irradiation compared to electron irradiation, possibly due to the larger molecular formation and destruction cross-sections in the case of the former. Interestingly, the column densities of CO and $C_2H_6$ were found to be equal after an administered radiation dose of approximately 63 eV per 16 u in both the proton and the electron irradiated ices.

Our results are of inherent interest to the astrochemistry occurring within icy grain mantles in dense interstellar clouds as they effectively demonstrate the production of complex organic species such as hydrocarbons and alcohols after processing by galactic cosmic rays, for which we have used energetic protons and electrons as an analogue. Such molecules may serve as feedstocks for the production of even more complex molecules of inherent interest to biology and biochemistry. Many of the daughter molecules produced by the irradiation of the $CH_4$:$H_2O$ ices may also be preserved and seeded to nascent planetary systems, as evidenced by their detections in comets and other Solar System objects. Our results therefore contribute to a better understanding and quantification of daughter molecules that can be produced as a result of different irradiation regimes that may dominate in different astronomical settings.

**Author Contributions:** Conceptualization, Duncan V. Mifsud; Data curation, Zuzana Kaňuchová; Formal analysis, Zuzana Kaňuchová; Funding acquisition, Nigel J. Mason; Investigation, Duncan V. Mifsud, Péter Herczku and Zuzana Kaňuchová; Methodology, Duncan V. Mifsud, Péter Herczku, Béla Sulik, Zoltán Juhász and Zuzana Kaňuchová; Resources, István Vajda and István Rajta; Supervision, Nigel J. Mason; Validation, Giovanni Strazzulla; Visualization, Sergio Ioppolo and Giovanni Strazzulla; Writing – original draft, Duncan V. Mifsud; Writing – review & editing, Duncan V. Mifsud, Péter Herczku, Béla Sulik, Zoltán Juhász, István Vajda, István Rajta, Sergio Ioppolo, Nigel J. Mason, Giovanni Strazzulla and Zuzana Kaňuchová.

**Funding:** The authors acknowledge support from the Europlanet 2024 RI which has received funding from the European Union's Horizon 2020 Research Innovation Program under grant agreement No. 871149. The main components of the ICA set-up were purchased using funds from the Royal Society obtained through grants Nos. UF130409, RGF/EA/180306, and URF/R/191018. Further developments of the installation were supported in part by the Eötvös Loránd Research Network through grants ELKH IF-2/2019 and ELKH IF-5/2020. This work has also received support from the European Union and the State of Hungary; co-financed by the European Regional Development Fund through grant No. GINOP-2.3.3-15-2016-00005. Support has also been received from the Research, Development, and Innovation Fund of Hungary through grant No. K128621. Duncan V. Mifsud is the grateful recipient of a University of Kent Vice-Chancellor's Research Scholarship. Zoltán Juhász acknowledges support from the Hungarian Academy of Sciences through the János Bolyai Research Scholarship. Sergio Ioppolo acknowledges support from the Danish National Research Foundation through the Centre of Excellence 'InterCat' (grant agreement No. DNRF150) and from the Royal Society. The research of Zuzana Kaňuchová is supported by the Slovak Grant Agency for Science (grant No. 2/0059/22) and the Slovak Research and Development Agency (contract No. APVV-19-0072).

**Data Availability Statement:** Data will be made available to interested parties upon reasonable request of one of the corresponding authors.

**Acknowledgments:** The authors would like to express their gratitude to Béla Paripás (University of Miskolc, Hungary) for his continued support and assistance, and to Thomas A. Field (Queen's University Belfast, United Kingdom) for his helpful comments and suggestions.

**Conflicts of Interest:** The authors declare no conflict of interest. The funders had no role in the design of the study; in the collection, analyses, or interpretation of data; in the writing of the manuscript; or in the decision to publish the results.